\documentclass[journal,12pt,onecolumn,draftclsnofoot,]{IEEEtran}
\usepackage{cite}

\ifCLASSINFOpdf
   \usepackage[pdftex]{graphicx}
\else
   \usepackage[dvips]{graphicx}
\fi

\usepackage{amsmath}
\usepackage{amssymb}

\ifCLASSOPTIONcompsoc
  \usepackage[caption=false,font=normalsize,labelfont=sf,textfont=sf]{subfig}
\else
  \usepackage[caption=false,font=footnotesize]{subfig}
\fi

\usepackage{stfloats}
\usepackage{url}

\usepackage{siunitx}
\DeclareSIUnit\sq{\ensuremath{\Box}}

\usepackage{xcolor}
\usepackage[normalem]{ulem}

\usepackage{chemformula}

\begin{document}

\title{Characterization of a photon-number resolving SNSPD using Poissonian and sub-Poissonian light}

\author{Ekkehart~Schmidt, 
				Eric~Reutter,
				Mario~Schwartz,
				H\"useyin~Vural,
				Konstantin~Ilin,
				Michael~Jetter,
				Peter~Michler,
				and~Michael~Siegel% <-this % stops a space
\thanks{This work was supported in part by DFG project SI704/10-1}
\thanks{E. Schmidt, K. Ilin and M. Siegel are with Institute of Micro- and Nanoelectronic Systems (IMS), Karlsruhe Institute of Technology, Hertzstrasse 16, Karlsruhe, Germany (email: ekkehart.schmidt@kit.edu)}% <-this % stops a space
\thanks{E. Reutter was with Institute of Micro- and Nanoelectronic Systems (IMS), Karlsruhe Institute of Technology, Hertzstrasse 16, Karlsruhe, Germany and is with Max Planck Institute for Solid State Research, Heisenbergstra{\ss}e 1, Stuttgart, Germany}% <-this % stops a space
\thanks{M. Schwartz, H. Vural, M. Jetter and P. Michler are with Institut f\"ur Halbleiteroptik und Funktionelle Grenzfl\"achen (IHFG), Center for Integrated Quantum Science and Technology (IQ\textsuperscript{ST}) and SCoPE, University of Stuttgart, Allmandring 3, Stuttgart, Germany}% <-this % stops a space
\thanks{Manuscript received ...; revised ...}
}

%\markboth{IEEE TRANSACTIONS ON APPLIED SUPERCONDUCTIVITY}%
\markboth{2EP\MakeLowercase{o}2C-10}{}%
%\markboth{\MakeLowercase{ar}X\MakeLowercase{iv}-preprint}{}%

\maketitle

\begin{abstract}
Photon-number resolving (PNR) single-photon detectors are of interest for a wide range of applications in the emerging field of  photon based quantum technologies. Especially photonic integrated circuits will pave the way for a high complexity and ease of use of quantum photonics. Superconducting nanowire single-photon detectors (SNSPDs) are of special interest since they combine a high detection efficiency and a high timing accuracy with a high count rate and they can be configured as PNR-SNSPDs. Here, we present a PNR-SNSPD with a four photon resolution suitable for waveguide integration operating at a temperature of 4 K. A high statistical accuracy for the photon number is achieved for a Poissonian light source at a photon flux below \num{5} photons/pulse with a detection efficiency of \SI{22.7(30)}{\percent} at \SI{900}{\nano\meter} and a pulse rate frequency of \SI{76}{\mega\hertz}. We demonstrate the ability of such a detector to discriminate a sub-Poissonian from a Poissonian light source.

\end{abstract}

\begin{IEEEkeywords}
Superconducting photodetectors, SNSPD, Photon number resolution, Nanowire, NbN.
\end{IEEEkeywords}

\section{Introduction}

\IEEEPARstart{P}{hoton}-number resolving (PNR) single-photon detectors allow the direct characterization of photon number (PN) states of an arbitrary photon source \cite{Schmidt.2018b} and the direct measurement of a photon-number correlation which can be used for quantum enhanced metrology \cite{Meda.2017} or for quantum optical communication \cite{Becerra.2015}. Random numbers for quantum cryptography can be generated using a PNR detector \cite{Applegate.2015}. PNR detectors with \num{4}-photon resolution are a necessary element for linear-optical quantum computation (LOQC) as proposed by Knill, Laflamme, and Milburn \cite{Knill.2001}. Integrated quantum photonic circuits are the next natural step to increase the complexity and ease of use of quantum photonic circuits \cite{OBrien.2009}. The realization of a fully integrated Hanbury Brown and Twiss (HBT) \cite{BROWN.1956} experiment on the single-photon level has recently been demonstrated on GaAs \cite{Schwartz.2018} as logic photonic building block. This allows the investigation of the brightness and the normalized 2nd order correlation (two-photon correlation) of the integrated single-photon source \cite{Glauber.1963} by the use of a \num{50}/\num{50} beam splitter in combination with two single-photon detectors. Furthermore this enables the on chip investigation of the single-photon character of emission. In principle the same results can be achieved using just a waveguide in combination with a PNR detector with at least \num{2}-photon resolution, recording two-photon detection events. This allows for a significant reduction in the circuit complexity of a quantum photonic circuit and simplifies the measurement: the multi-photon character has not to be evaluated in a time-correlated measurement but can be measured directly. The combination of two \num{2}-photon resolving detectors with two splitters and a delay line is enough for the simultaneous full characterization of a single-photon state including its indistinguishably \cite{Thomay.2017}. This shows that PNR-detectors allow for a reduction of the number of circuit elements and  will simplify the fabrication and use of a quantum photonic integrated circuit. An ideal detector for quantum photonics should have a high detection efficiency, a good timing resolution, sufficient count rate and should be easy to fabricate and operate.
Superconducting nanowire single-photon detectors (SNSPDs) can be designed to have close to unity detection efficiency \cite{Lita.2008,Baek.2011,Marsili.2013} and can be fabricated out of a single superconducting layer. SNSPDs can reach up to \si{\giga\hertz} count rates and timing jitters in the \si{\pico\second} range \cite{Korzh.4182018,Sidorova.2017}, but are click/no-click detectors without intrinsic energy and PNR. PNR capability for SNSPDs can be achieved by amplitude multiplexing. In this case only one readout and biasing line for each detector is needed and they can be used with a standard SNSPD setup. Amplitude multiplexing can be realized using a concept proposed by Jahanmirinejad et al. in 2012 \cite{Jahanmirinejad.2012}, by dividing the detector into several sections by placing resistors in parallel to series nanowires, resulting in an output pulse with an amplitude related to the number of triggered elements. The concept was proven for \ch{NbN}-SNSPDs on \ch{GaAs}\cite{Jahanmirinejad.2012b} and has been extensively studied \cite{Mattioli.2014,Zhou.2014,Mattioli.2015}. PNR-SNSPDs based on series nanowires have been shown to be able to operate over a high dynamic range of photons \cite{Zhou.2014,Mattioli.2015}. Furthermore their waveguide integration on \ch{GaAs} has been demonstrated \cite{Sahin.2013}. 
However, all demonstrations of series PNR-SNSPDs so far were performed in fiber coupled setups at temperatures below \SI{2.1}{\kelvin} using a laser as a photon source at a wavelength of \SI{1310}{\nano\meter} \cite{Jahanmirinejad.2012,Jahanmirinejad.2012b,Mattioli.2014,Zhou.2014,Mattioli.2015}. Free space accessible helium flow cryostats operating at \SI{4}{\kelvin} are widely used in the photonic community for excitation and characterization of quantum dot (QD) single-photon sources. They allow easier access to the sample and fast modifications of the optical setup, which are key to employ complex QD excitation schemes. The demonstration of a series PNR-SNSPD at \SI{4}{\kelvin} in a free space cryostat and a direct demonstration of the response of such a detector to a QD single-photon source so far remains elusive.\\
\indent In this work, we present a series PNR-SNSPD in a design optimized for waveguide integration (II). The detector is characterized using two photon sources with well known statistics: a laser source and a QD-single photon source. The detector is illuminated from the top by external photon sources. Top illumination allows for a well controlled, spatial homogeneous, photon flux on the detector area in contrast to waveguide coupling \cite{Sahin.2013} and also the use of a resonantly excited QD single-photon source without introducing stray light \cite{Schwartz.2018}. This ensures an accurate characterization of the optical properties of the detector. We prove the single-photon operation of the device at \SI{4}{\kelvin} in a free space cryostat and demonstrate the PNR capability at $\lambda=\SI{900}{\nano\meter}$ utilizing a pulsed laser source. Finally, we demonstrate the capability of our PNR-SNSPD to reveal the sub-Poissonian statistic of a QD light source.

\section{Design and fabrication}

\begin{figure}[!t]
\centering
\subfloat[]{\includegraphics[width=1.6in]{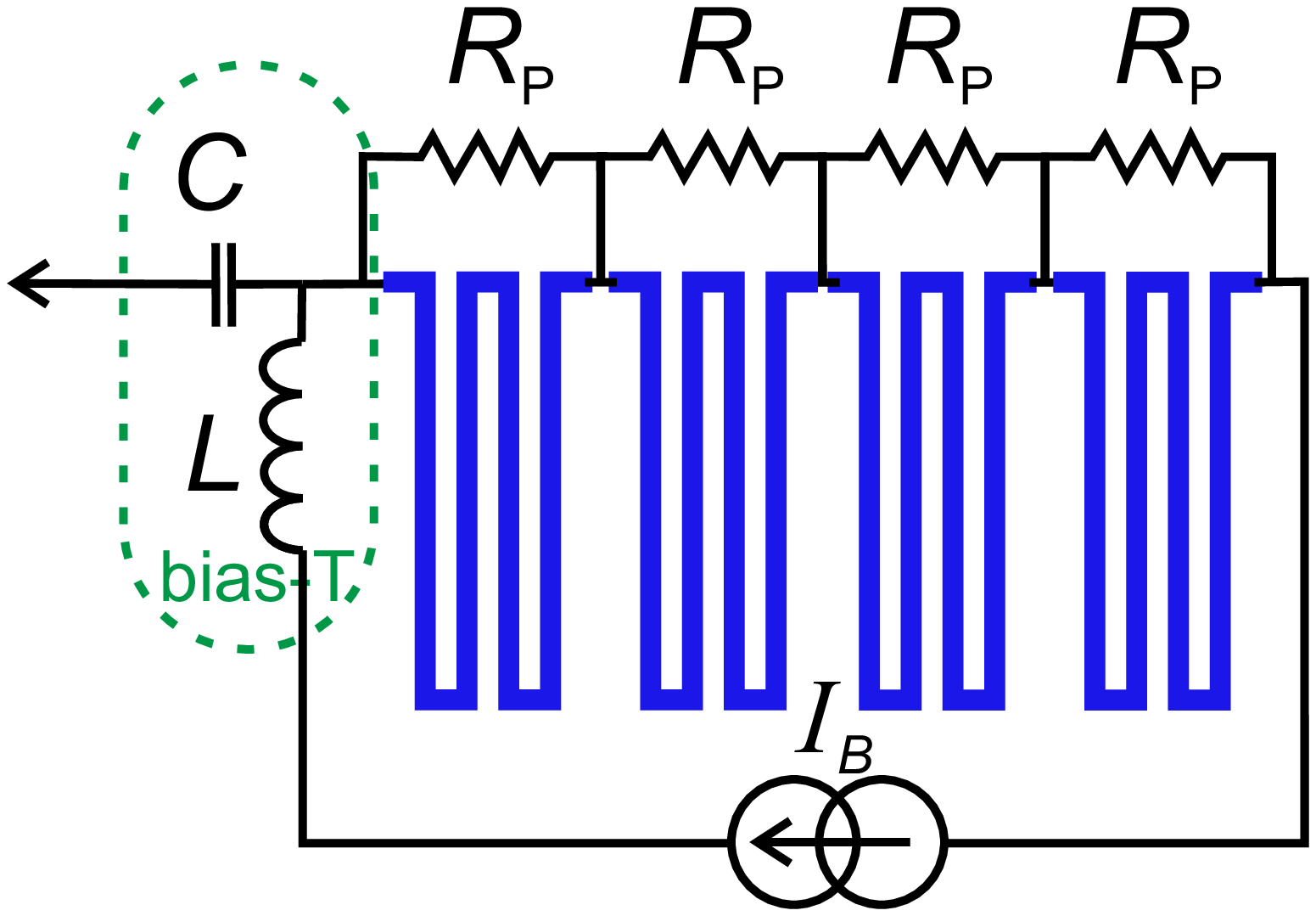}%
\label{Figure1a}}
\hfil
\subfloat[]{\includegraphics[width=1.6in]{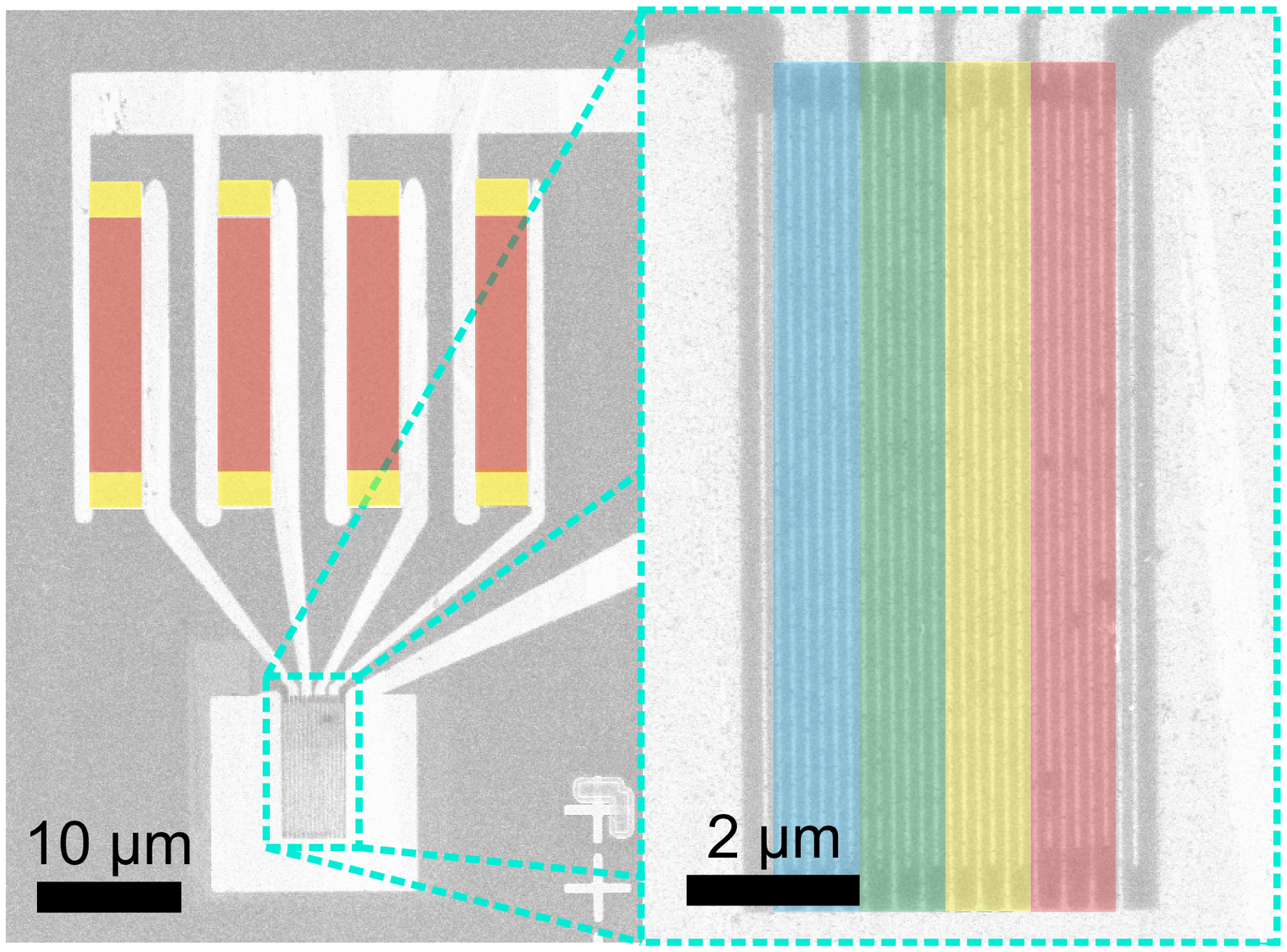}%
\label{Figure1b}}
\caption{(a) Electrical representation of a \num{4}-pixel series PNR-SNSPD. (The nanowire is colorized blue.) \newline (b) SEM image of a \num{4}-pixel PNR-SNSPD. Left: overview image with resistors colorized in red and contact pads in yellow. Right: close up of the nanowire, pixels are colorized individually.}
\label{Figure1}
\end{figure}

A principle scheme of the electrical circuit of a four-pixel series PNR-SNSPD is shown in Figure \ref{Figure1a}. The photon detection mechanism of PNR-SNSPDs is identical to standard single-pixel SNSPDs: an absorbed photon causes the formation of a resistive region across the nanowire with assistance of an applied current, $I_\textrm{bias}$, slightly smaller than the critical current $I_\textrm{C}$. However, in case of PNR-SNSPDs, the bias current is diverted from the ignited pixel to the adjacent parallel resistor, $R_\textrm{P}$, thus resulting in a voltage pulse $\approx I_\textrm{bias}\cdot R_\textrm{P}$. While, after the detection event, the triggered pixel remains insensitive until it returns into the superconducting state, all other pixels are fully biased and still able to detect photons. If several photons ($N_\textrm{ph}$) are absorbed in the detector simultaneously, each in different pixels, the current is redistributed to the correspondent resistors and thus an output pulse is proportional to the number of absorbed photons, $\approx N_\textrm{ph}\cdot I_\textrm{bias}\cdot R_\textrm{P}$.
The detector (Figure \ref{Figure1b}) was specifically designed for integration on top of a tapered waveguide as shown in \cite{Schwartz.2018}. To allow the detectors to sit on the waveguide and the resistors to face away from the waveguide, the resistors were placed to one side and the nanowires to the other side. In addition, this enables an easy adjustment of the photon absorption by changing the length of the nanowire on the waveguide. To allow \SI{4}{\kelvin} operation, the SNSPD is made from NbN thin film. GaAs was chosen as a substrate, since it allows for the monolithic integration of all necessary photonic elements for an integrated quantum photonic chip \cite{Schwartz.2018}. The growth of high quality NbN films on GaAs is challenging and is promoted by the AlN buffer layer. The detector was patterned out of a \SI{4.5}{\nano\meter} thick NbN film (zero-resistance critical temperature of as-deposited film was $T_\textrm{C} = \SI{10.3}{\kelvin}$) deposited on top of a \SI{10}{\nano\meter} thick AlN buffer layer using electron beam lithography and reactive-ion etching with \ch{SF_6} and \ch{O_2}. Details on the fabrication of NbN films and AlN buffer layers on GaAs substrates can be found in \cite{Schmidt.2017}. The nanowire is patterned to a width of \SI{105(5)}{\nano\meter} and a length of \SI{60}{\micro\meter} for each individual pixel. The full detector has \num{4} pixels and an active area of \SI{4 x 10}{\micro\meter}. The resistors, $R_\textrm{P}$, are made from DC-magnetron sputtered palladium and patterned using a lift-off process. They have a thickness of \SI{30}{\nano\meter} and a sheet resistance of \SI{5}{\ohm\per\sq} at \SI{4.2}{\kelvin}. With a length corresponding to five squares, each resistor is designed to have a resistance of \SI{25}{\ohm}. To ensure a low contact resistance to the NbN, the Pd is sputtered on top of the contact pads made from an Nb/Au bilayer, with a thickness of \SI{9}{\nano\meter}/\SI{14}{\nano\meter}. 

\section{Superconducting properties}

\begin{figure*}[!t]
\centering
\subfloat[]{\includegraphics[width=2.1in]{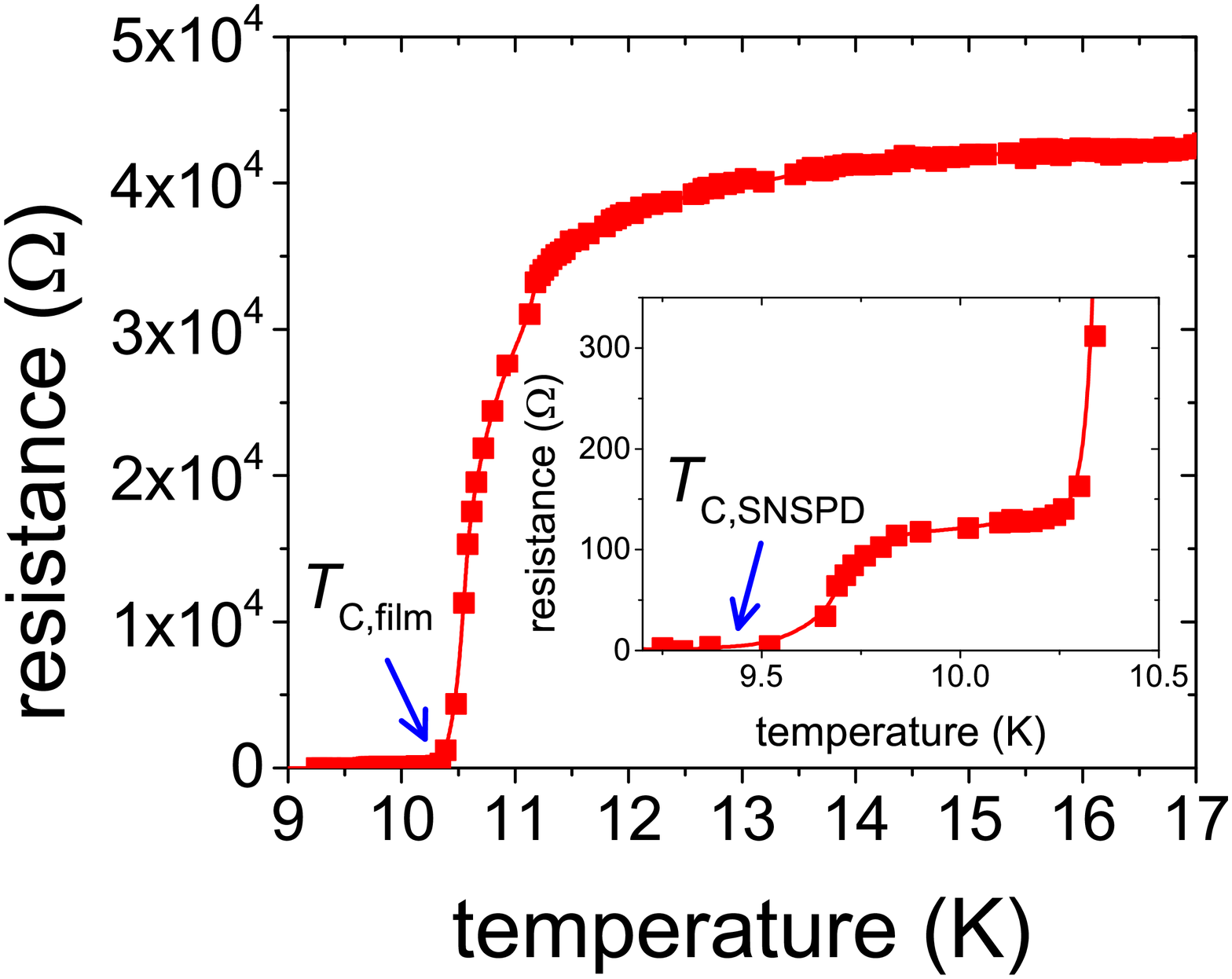}%
\label{Figure2a}}
\hfil
\subfloat[]{\includegraphics[width=1.9in]{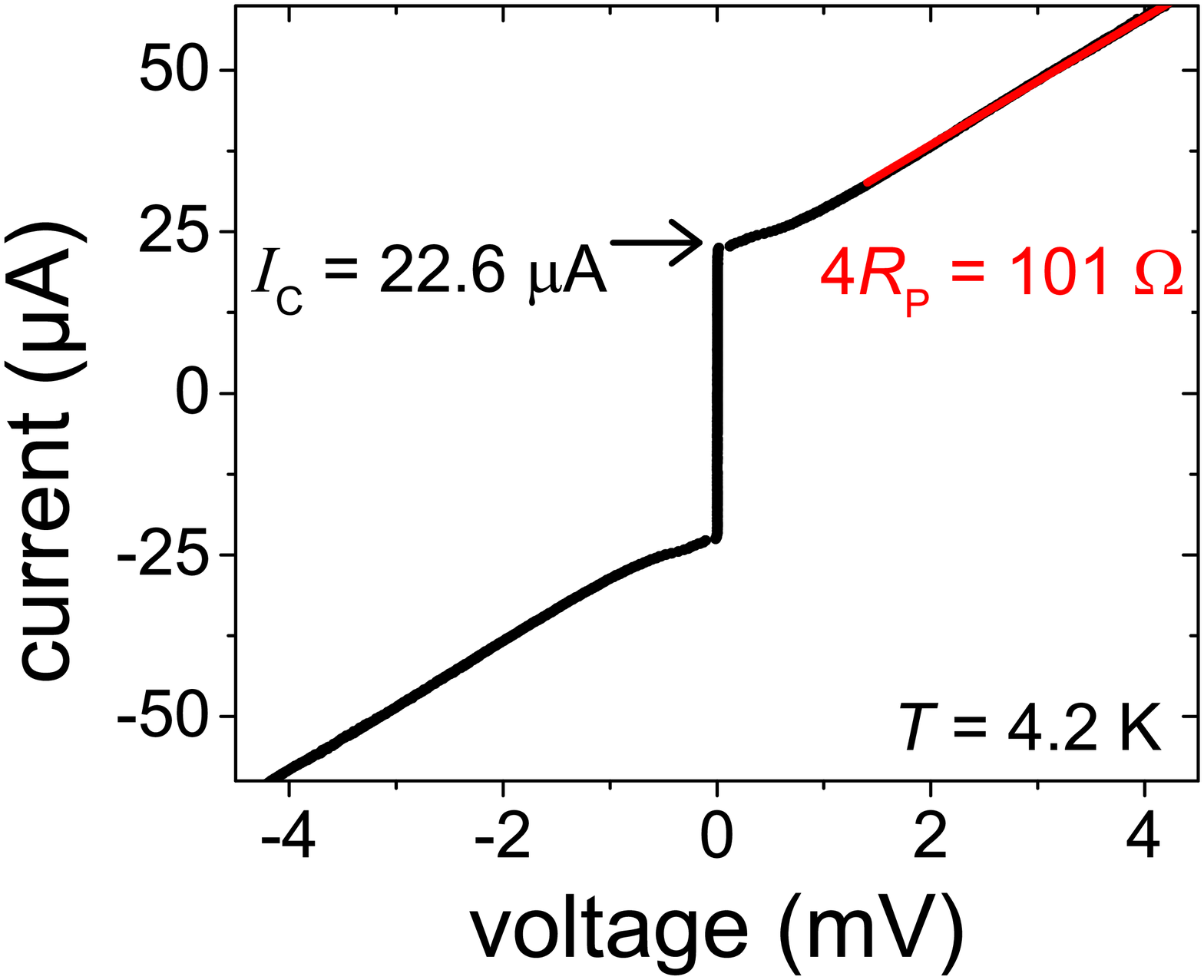}%
\label{Figure2b}}
\hfil
\subfloat[]{\includegraphics[width=2.1in]{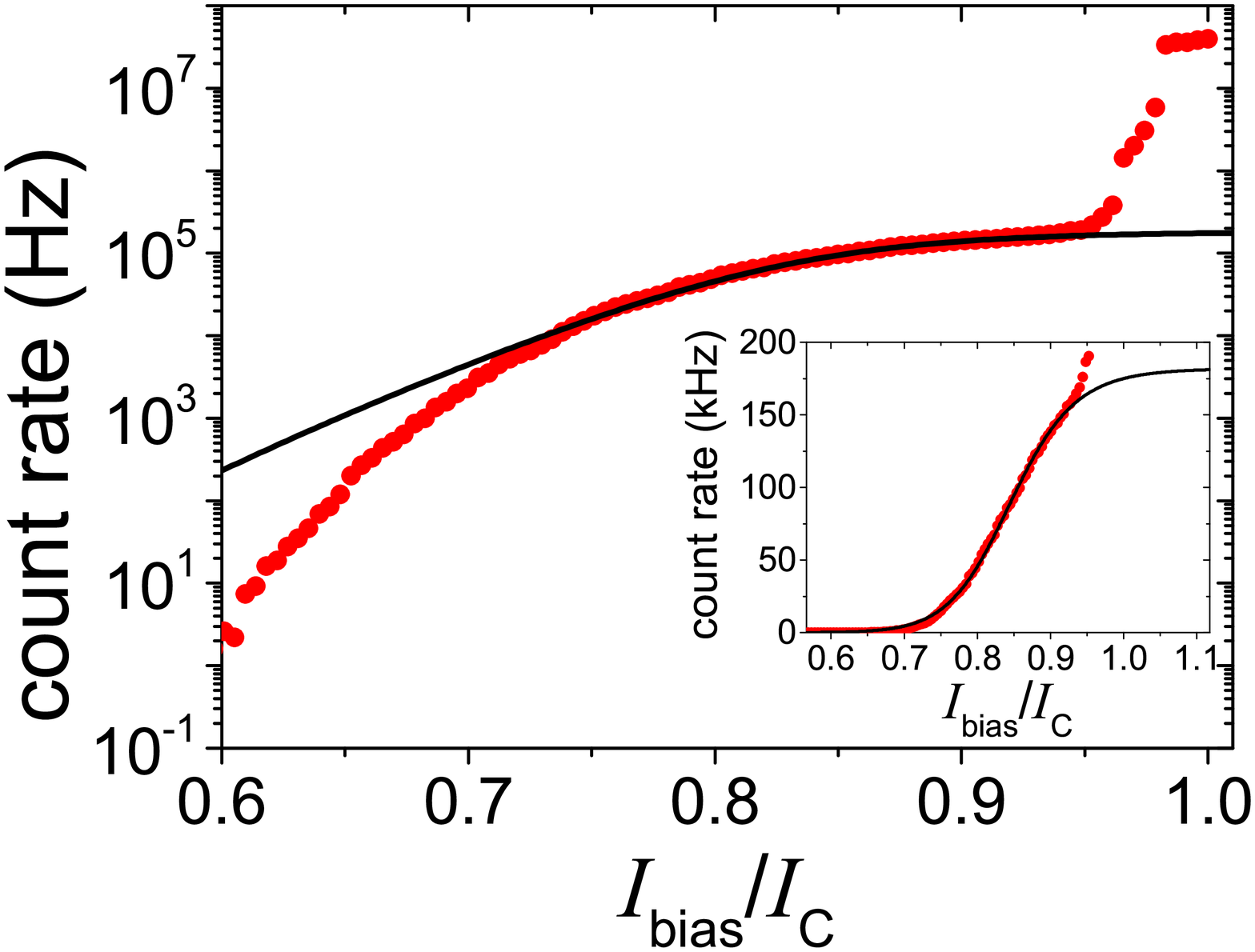}%
\label{Figure2c}}
\caption{(a) $R(T)$-characteristic of the detector. The inset is a close up at the foot of the superconducting transition. (b) $I(V)$-characteristic of the detector. The value of $R_\textrm{P}$ is extracted from a fit to the linear region.  (c) Count rate dependence as a function of the applied bias current. The inset depics the count rate in linear scale. The black line is a logistic sigmodial fit to the data $<0.93 I_\textrm{C}$ to extract the saturation level.}
\label{Figure2}
\end{figure*}

The $R(T)$-dependence of the detector reveals a two-step superconducting transition (Figure \ref{Figure2a}). The first step corresponds to the transition of the micron sized coplanar readout line made from the same NbN film into the superconducting state and reveals a critical temperature $T_{\textrm{C}} = \SI{10.3}{\kelvin}$ which is equal to $T_\textrm{C}$ of the as-deposited NbN film. The second step in the $R(T)$-curve seen at $T < \SI{10}{\kelvin}$ (see inset in Fig. \ref{Figure2a}) corresponds to the transition of the nanowire with zero-resistance $T_\textrm{C} = \SI{9.4}{\kelvin}$. The resistance about \SI{120}{\ohm} seen just before the second superconducting transition is mostly determined by the resistance of the four series resistors shunting the NbN nanowire with $R_\textrm{SNSPD} >> 4R_\textrm{P}$ at \SI{10}{\kelvin}. The current-voltage characteristic ($IV$-curve) of the PNR-SNSPD measured at \SI{4.2}{\kelvin} (Fig. \ref{Figure2b}) demonstrates just a single-step transition into the resistive state at $I_\textrm{C} = \SI{22.6}{\micro\ampere}$ thus indicating a high homogeneity of detector pixels. The differential resistance of the device in the resistive state at currents well above the $I_\textrm{C}$ value is about \SI{100}{\ohm} and corresponds to the total resistance of the four series resistors, at \SI{4.2}{\kelvin} giving an individual resistance of each resistor of about \SI{25}{\ohm} which corresponds well to the design value.
For optical characterization the device was mounted inside a helium flow cryostat and kept at a temperature about \SI{4}{\kelvin}. The detector was excited from the top by photons passing through an optical window of the cryostat. The exciting photons at a wavelength $\lambda = \SI{900}{\nano\meter}$ (these photons are of energy typical for emission of resonantly excited In(Ga)As/GaAs quantum dots \cite{Schwartz.2018,Reithmaier.2013,Schwartz.2016,Vural.2018}) were selected from radiation of a thermal halogen light source using a monochromator with a spectral resolution of \SI{10}{\nano\meter}. The bias current dependence of count rate (Fig. \ref{Figure2c}) of the studied PNR-SNSPD was fitted with a sigmodial logistic function to extract the saturation level of the detector. The fit was done for $I_\textrm{bias} < 0.93I_\textrm{C}$, since for $I_\textrm{bias} > 0.93I_\textrm{C}$ the count rate is dominated by dark counts and relaxation oscillations \cite{Kerman.2009}. Out of the fit, the intrinsic detection efficiency at a bias current of $0.90I_\textrm{C}$ can be estimated to \SI{77}{\percent}\cite{Vodolazov.2015}.

\section{Photon number resolving}

\begin{figure*}[!t]
\centering
\subfloat[]{\includegraphics[width=2.1in]{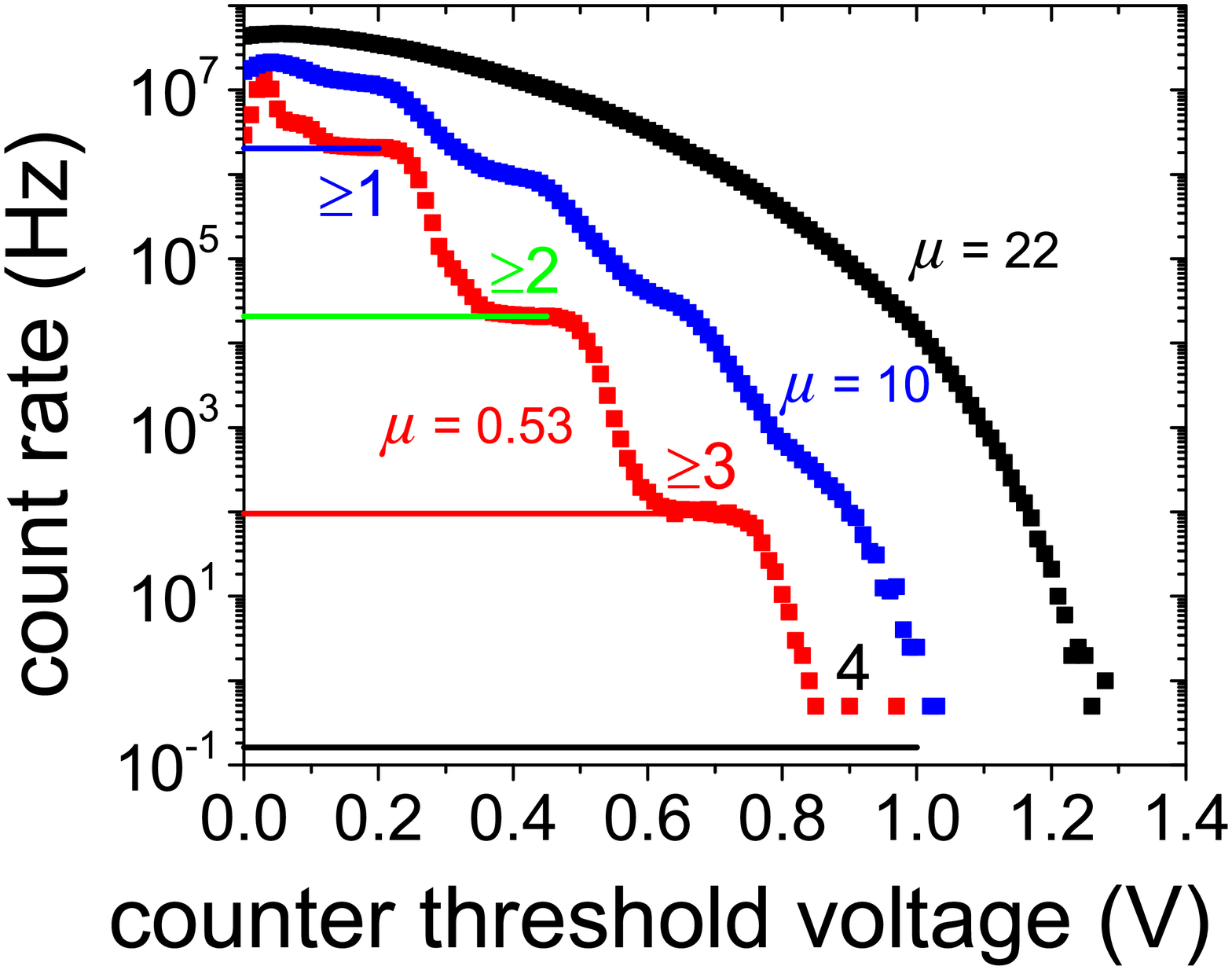}%
\label{Figure3a}}
\hfil
\subfloat[]{\includegraphics[width=2in]{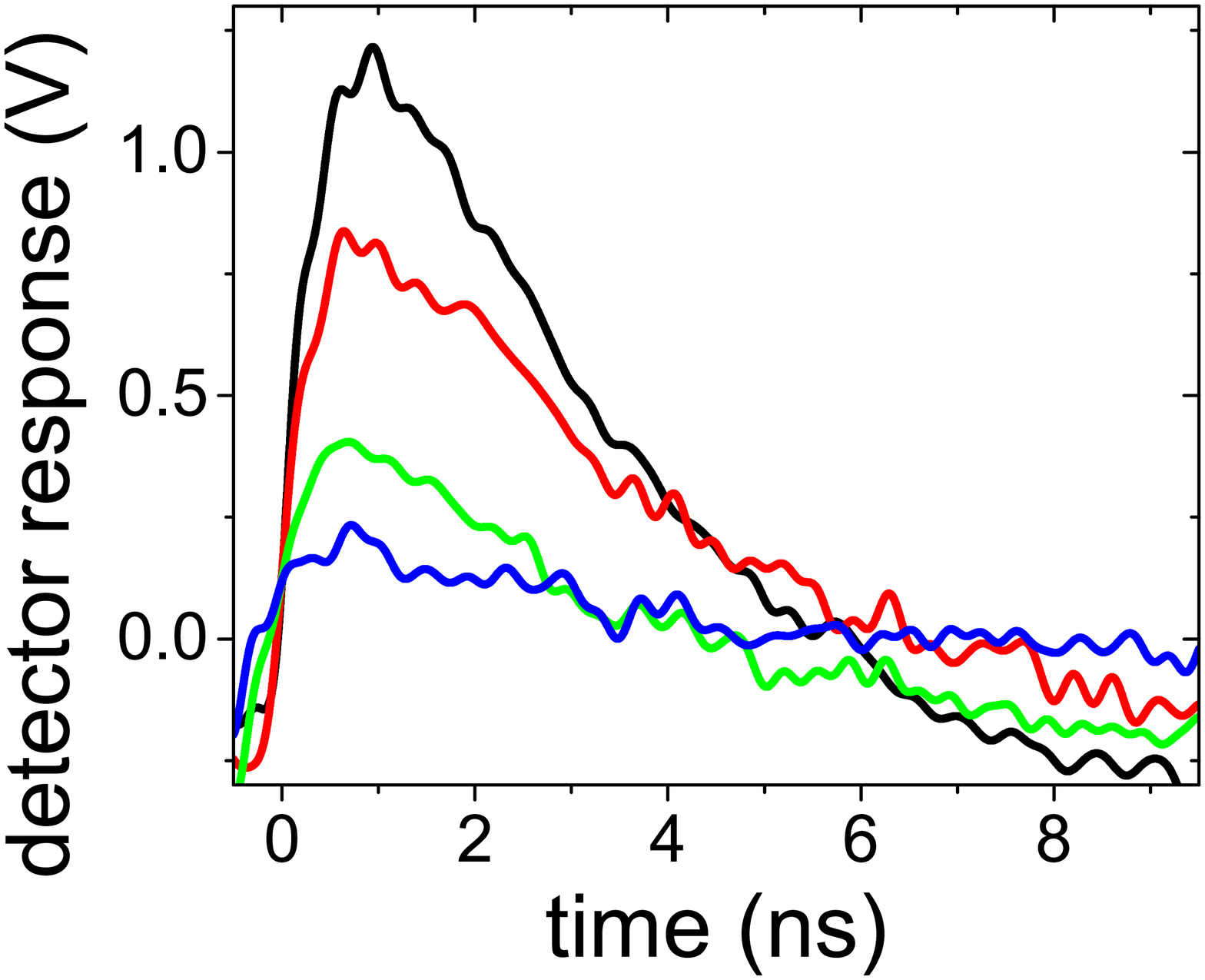}%
\label{Figure3b}}
\hfil
\subfloat[]{\includegraphics[width=2.15in]{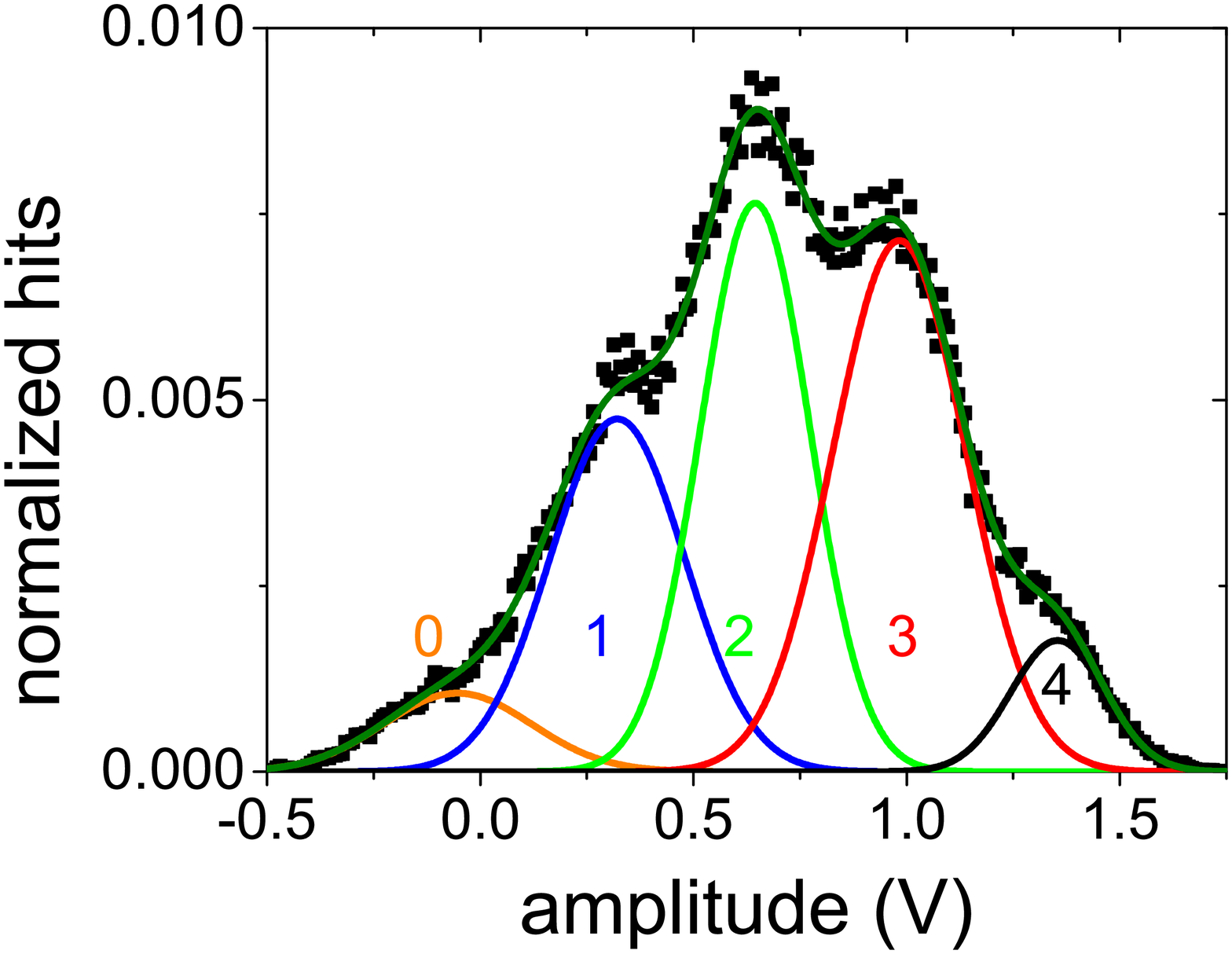}%
\label{Figure3c}}
\caption{(a) Count-rate dependence on the counter threshold voltage for different mean numbers of photons per pulse $\mu$. Numbers inside the graph correspond to the individual photon levels, the solid lines correspond to the expected photon levels out of equation (\ref{Fit}).
(b) Individual pulses at different photon levels for $\mu>20$.
(c) Histogram of the maximal amplitude of detector pulses for one million recorded pulses at $\mu = 42$. The histogram is normalized to unity. The solid lines are Gaussian fits to the individual detector photon levels and the sum of all Gaussian fits (dark green).}
\label{Figure3}
\end{figure*}

The PNR capability was investigated with a Poissonian light source: a pulsed Ti:Sapphire laser at a wavelength of \SI{900}{\nano\meter}, pulse length of \SI{3.6}{\pico\second} and a pulse repetition frequency (PRF) of \SI{76}{\mega\hertz} in combination with a variable attenuator. The detector was biased at $0.9I_\textrm{C}$ to ensure stable and low noise operation (Figure \ref{Figure2c}). The count rate was investigated in dependence on the threshold voltage of the counting electronic at a mean number of photons per pulse $\mu = 0.53$ (Figure \ref{Figure3a} red). All rising edges that exceed the specific threshold voltage are counted. The dependence shows a noise peak at \SI{50}{\milli\volt}, followed by four flat equidistant regions ($\Delta V \approx\SI{250}{\milli\volt}$). The existence of four flat regions demonstrates the operation of all pixels and four photon levels of the detector: The flat region \num{1} (blue) corresponds to the detection of at least one ($\geq 1$), region \num{2} (green) of $\geq 2$, region 3 (red) of $\geq 3$ and finally region \num{4} (black) of four photons. Consequently the photon count rate for each PN can be conveniently measured with a pulse counter by triggering to the corresponding flat region and subtracting the count rate of the next higher amplitude region. For an increasing $\mu$, the amplitude levels start to shift and overlap ($\mu = 10$) until no flat regions are visible any more ($\mu = 22$). 
Example pulses for a large $\mu$ recorded at the four individual photon levels are depicted in Figure \ref{Figure3b}. The pulses have a $1/e$ decay time of $\tau_{1/e} = \SI{4.4}{\nano\second}$. The base level of individual pulses changes in dependence on the amplitude of the previous pulse. This is caused by a charging of the readout line caused by the increasing number of high amplitude pulses for an increasing $\mu$ and the high PRF of \SI{76}{\mega\hertz}. This change of ground potential contributes to the amplitude jitter of detector pulses and causes an overlap of amplitude levels at large $\mu$.
To evaluate the PN distribution for a high $\mu$, the pulse amplitudes were recorded in a histogram using a \SI{16}{\giga\hertz} real time oscilloscope with a trigger signal from the laser (Figure \ref{Figure3c}). The histogram is fitted with a sum of Gaussian distributions (dark green) with a Gaussian for each amplitude level of the detector including the \num{0}-level (no detected photons: orange). The relative area of the Gaussian for each amplitude level normalized on the area of the full histogram gives the recorded PN distribution. We recorded PN distributions for $\mu$ from \num{0.35} to \num{144}. At small values of $\mu$ the recorded PN distributions consist mainly of 0-photons or 1-photon events. With an increase of $\mu$ the probability for higher PN detector responses increases: the experimental PN response of the detector is proportional to the input power. To evaluate the fidelity of our detector we compared the measured PN distribution with the theoretically expected photon statistic. The probability $P_\eta^N(n|\mu)$ of detecting $n$ photons from an optical pulse out of a light source with a Poissonian statistic using an $N$-element detector with detection efficiency $\eta$ is described by \cite{Fitch.2003}\cite{Dauler.2009}:

\begin{align}
	\begin{split}
		P_\eta^N(n|\mu) =\sum^{\infty}_{m=n}{\frac{N!}{n!(N-n)!}\frac{(\eta \mu)^m e^{-\eta \mu}}{m!}}\times \\ \sum^n_{j=0}{(-1)^j \frac{n!}{j!(n-j)!}\left[1-\eta+\frac{(n-j)\eta}{N}\right]^m}
	  \label{Fit}
	\end{split}
\end{align}

\noindent This statistic is only valid for a uniform detector with an identical efficiency of all pixels. We fitted our measured experimental PN distribution with the theoretical distribution $P_\eta^N(n|\mu)$ using $\eta$ as a free fit parameter.

\begin{figure*}[!t]
	\centering
	\subfloat[]{\includegraphics[width=2.5in]{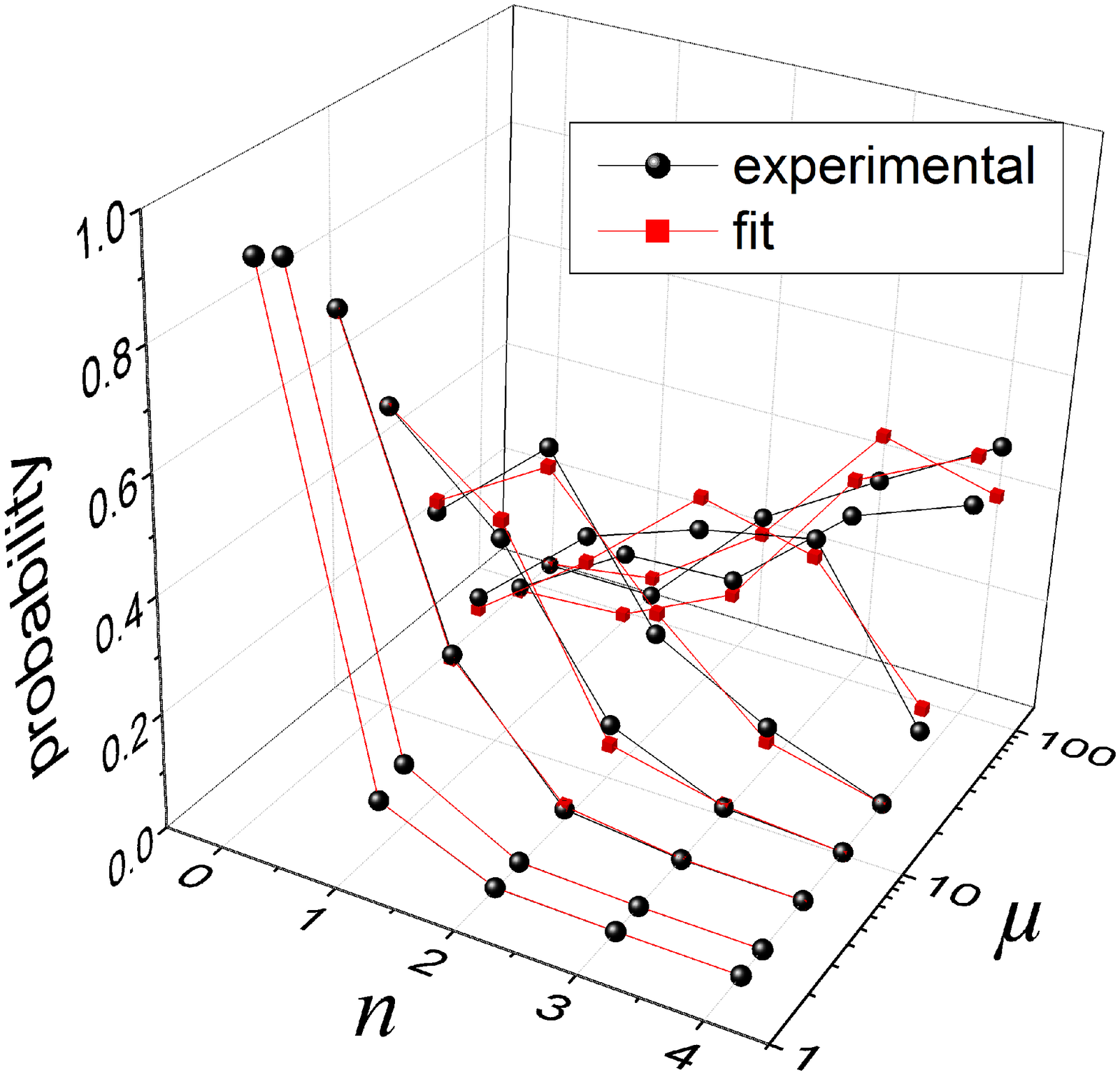}%
	\label{Figure4a}}
	\hfil
	\subfloat[]{\includegraphics[width=2.1in]{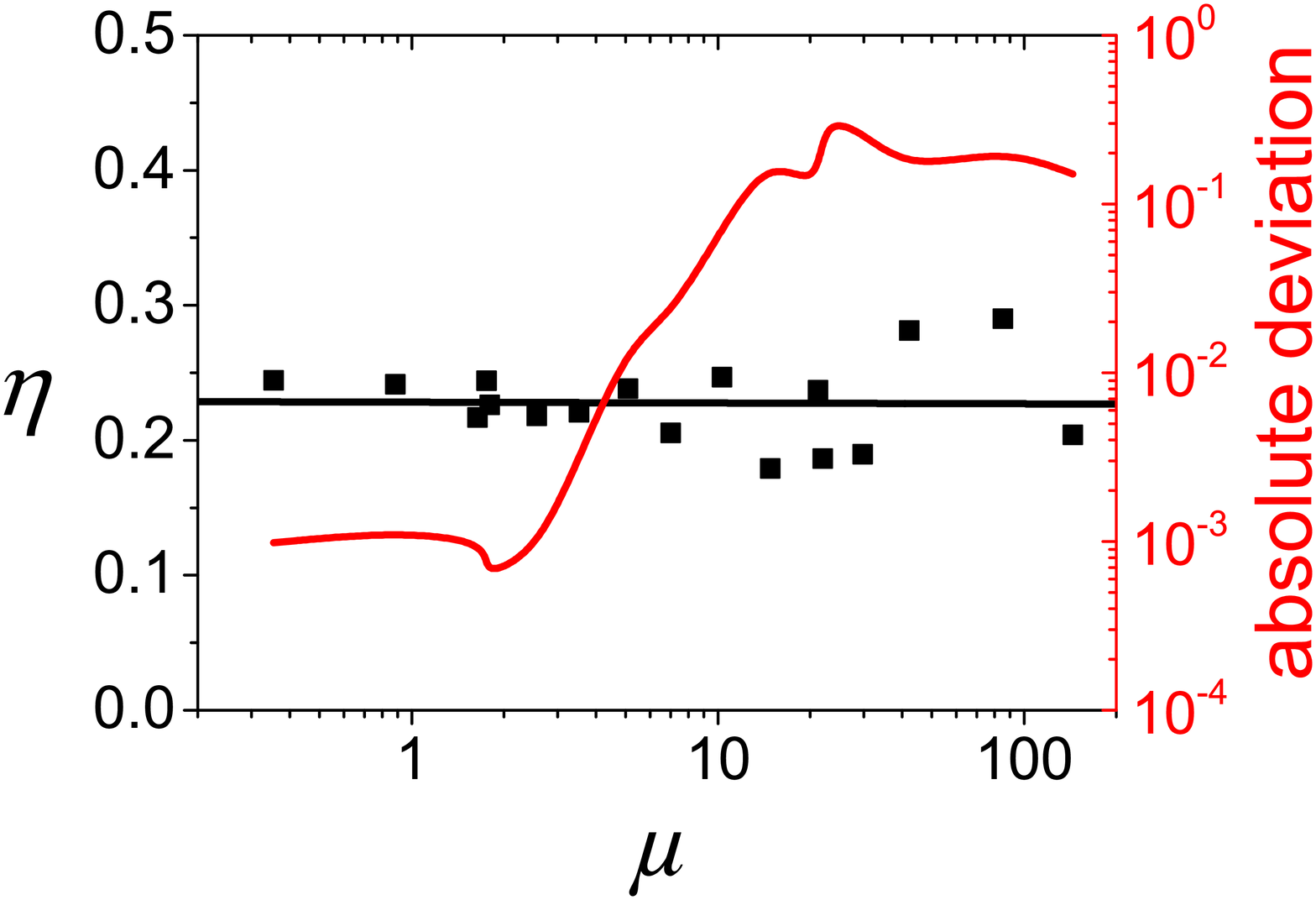}%
	\label{Figure4b}}
	\caption{(a) Comparison of the experimental PN distributions at different $\mu$ with the fit of the theoretical distribution (\ref{Fit}). Not all distributions are shown for visibility. (b) The detection efficiency $\eta$ from the theoretical fit in relation to the number of photons/pulse ($\mu$). The black solid line is a linear fit to $\eta$. The deviation of the fit from the measured distribution is shown in red.}
	\label{Figure4}
\end{figure*}

The fit closely approximates the experimental results by the theoretical distribution for a small $\mu$ (Figure \ref{Figure4a}). For an increasing $\mu$ a deviation of the fit from the experimental data is visible. All $\eta$ extracted out of the fit of $P_\eta^N(n|\mu)$ for each $\mu$ are depicted in Figure \ref{Figure4b}. The absolute deviation shown in the same graph is the sum of the deviations of the experimental distribution from the fit for each PN. For $\mu$ smaller than \num{5}, the fitted efficiency has only a small scatter and the absolute deviation is less than \SI{1}{\percent}. This demonstrates that the detector statistic closely resembles the Poissonian statistic of the laser for a small $\mu$ and furthermore that the efficiency of all pixels is identical. For $\mu$ larger than \num{5}, $\eta$ shows a significant scatter and the absolute deviation from the fit is in the order of \SI{20}{\percent}. A linear fit to all extracted $\eta$ reveals a flat average $\eta$ over the full investigated photon range of $\mu$ at \SI{22.7(30)}{\percent} (Figure \ref{Figure4b}). This $\eta$ is in good agreement with the relative distribution of count rates for flat regions in Figure \ref{Figure3a}. A stable $\eta$ indicates a high homogeneity of detector pixels \cite{Zhou.2014} and that the count rate of the detector is sufficient to resolve the PN distribution of a light source at a PRF = \SI{76}{\mega\hertz} for a $\mu$ from \num{0.35} to \num{144} in contrast to the used readout. We assign the large scatter of $\eta$ and the deviation of the experimental distribution from the fit at a $\mu>\num{5}$ to our readout chain: the merging of amplitude levels caused by a lift of the base level due to piling of pulses at the readout for a large $\mu$, makes the Gaussian fit difficult and inaccurate for extracting the PN distribution. Nevertheless, we successfully demonstrated the accuracy of our measurement setup for $\mu \leq \num{5}$. This is sufficient because quantum photonic circuits usually operate on a single-photon level and for LOQC the required photon number is $\leq \num{4}$ \cite{Knill.2001}.

\section{Poissonian vs sub-Poissonian light source}

\begin{figure*}[!t]
\centering
\subfloat[]{\includegraphics[width=2.1in]{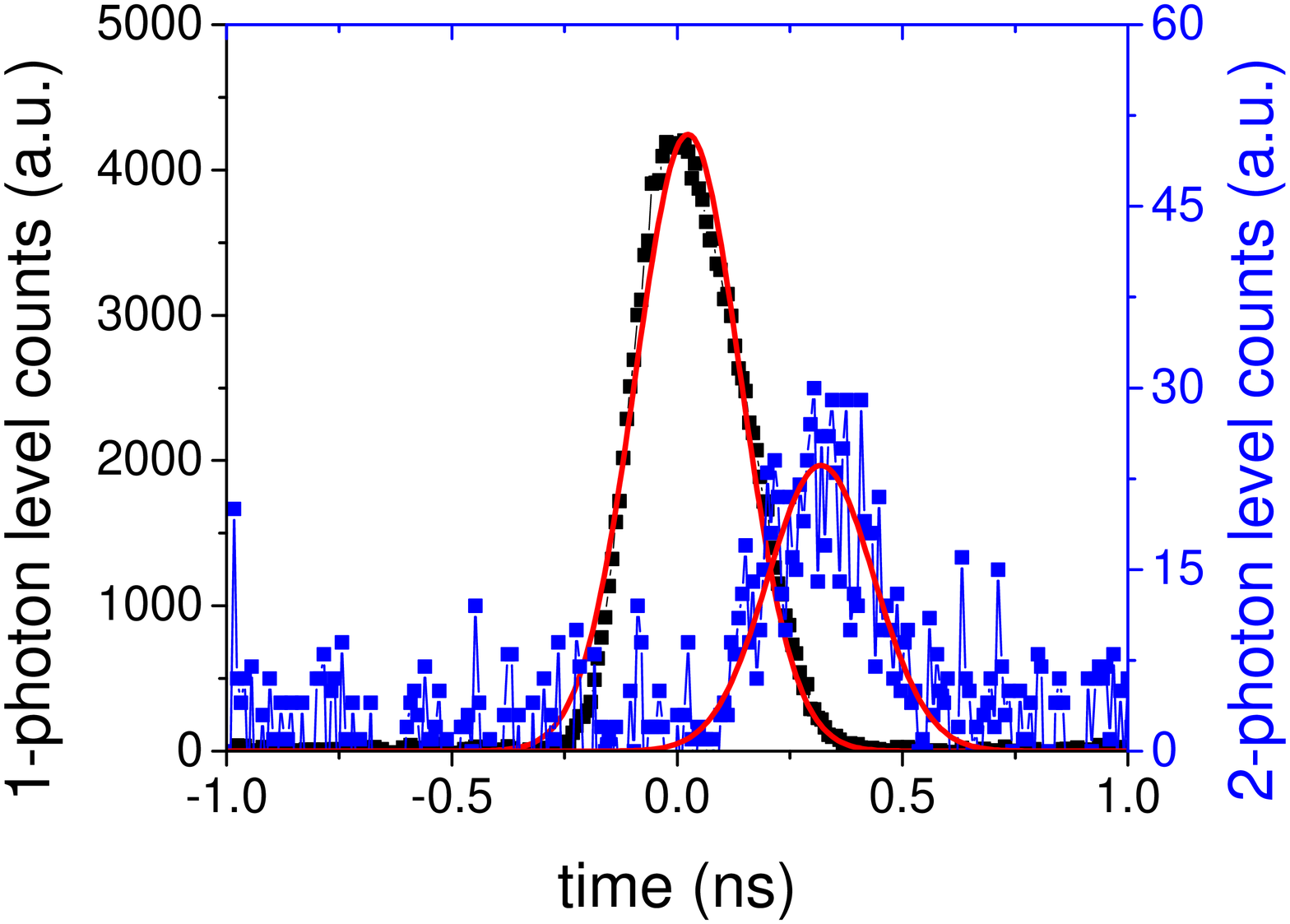}%
\label{Figure5a}}
\hfil
\subfloat[]{\includegraphics[width=2.1in]{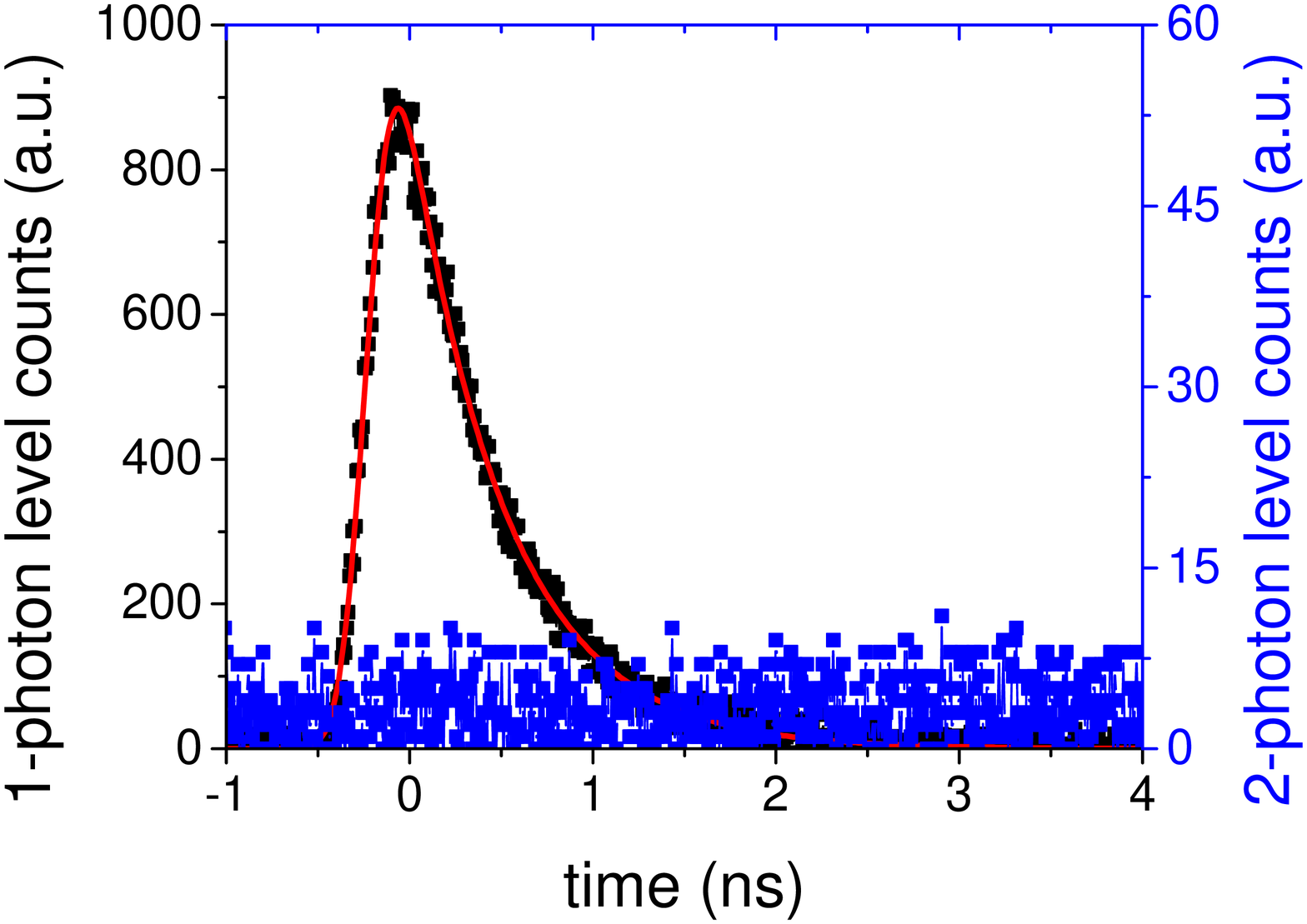}%
\label{Figure5b}}
\caption{\num{2}-photon level TCSPC measurement for a pulsed laser (a) and a resonantly excited quantum dot single-photon source (b). The \num{1}-photon level is displayed in black and the \num{2}-photon level in blue. The red solid lines are fits to the data.}
\label{Figure5}
\end{figure*}

We evaluated the capability of our PNR-SNSPD to distinguish a Poissonian from a sub-Poissonian light source. We compared the detector response to a pulsed laser (Poissonian light source) to the response to a pulsed QD (sub-Poissonian light source) on the \num{1}- and \num{2}-photon level. We performed this experiment in a time-correlated measurement to further resolve the temporal characteristic of the light source. The detector was illuminated through free space using a pulsed Ti:Sapphire laser and a resonantly excited In(Ga)As/GaAs QD. The QD is mounted and excited in a separate cryostat. Details on the used QD and its excitation can be found in the supplementary material of \cite{Vural.2018}. To provide the best comparability we adjusted the laser emission to a wavelength (\SI{854}{\nano\meter}) and photon flux ($\approx \num{1000}$ counts/second) comparable to the emission of the QD. The detector response was measured using a Picoquant HydraHarp TCSPC electronic using a trigger signal from the laser as time reference. The \num{1}- and \num{2}-photon levels of the detector where recorded simultaneously by using two input channels at different trigger levels. The measurement was performed at a bias level of $0.8I_\textrm{C}$ to increase the signal to noise ratio of our measurement. Prior to evaluation the average noise level was subtracted from the data. The detector response to laser illumination (Figure \ref{Figure5a}) shows a detection of photons on the \num{1}- as well as on the \num{2}-photon level. The \num{2}-photon level shows a small delay due to a slightly longer cable. The timing distributions on both photon levels correspond to the instrumental response function (IRF) of the setup. From a Gaussian fit, a standard deviation $\sigma = \SI{98(1)}{\pico\second}$ is retrieved for both photon levels. The detector response to QD photons (Figure \ref{Figure5b}) shows a detection of photons on the \num{1}- but only noise on the \num{2}-photon level. This in comparison to the laser clearly proves the sub-Poissonian character of the QD source. The timing distribution measured for the QD (Figure \ref{Figure5b}) was fitted with an exponentially modified Gaussian (EMG) distribution using the decay time of the dot as a free fit parameter. The EMG resembles the Gaussian distribution of the IRF convoluted with a monoexponential probability density function caused by the monoexponential decay of a QD. Out of the fit an IRF with $\sigma = \SI{114(1)}{\pico\second}$ and decay time of \SI{525(3)}{\pico\second} are extracted. The decay time is in good agreement with the decay of this QD measured with an avalanche photo-diode \cite{Vural.2018}. This proves the suitability of our detector for discriminating a sub-Poissonian from a Poissonian light source.

\section{Conclusion}

We demonstrated a photon-number resolving SNSPD suitable for waveguide integration on GaAs at a temperature of \SI{4}{\kelvin} in a free space accessible cryostat. PNR resolution was demonstrated for \num{4} photons at \SI{900}{\nano\meter} with a high statistical accuracy at photon rates below \num{5} photons/pulse with a detection efficiency of \SI{22.7(30)}{\percent} at a pulse rate frequency of \SI{76}{\mega\hertz}. This detection efficiency can be enhanced in future by the waveguide integration of our detector. The discrimination of a sub-Poissonian from a Poissonian light source was successfully demonstrated.

\section*{Acknowledgment}
\addcontentsline{toc}{section}{Acknowledgment}

We like to thank F. Hornung from the Institut f\"ur Halbleiteroptik und Funktionelle Grenzfl\"achen (IHFG), University of Stuttgart, for helpful discussions during the preparation of this manuscript.

\bibliographystyle{IEEEtran}
\bibliography{ASC2018}

\end{document}